\documentstyle[twocolumn,floats,prl,aps,epsf,psfrag]{revtex}
\begin{document}
\draft

\title{Quantum error correction of dephasing in 3 qubits}
\author{Samuel L.~Braunstein\cite{www}}
\address{Universit\"at Ulm, Abteilung Quantenphysik, 89069 Ulm, Germany}
\date{\today}
 
\maketitle

\begin{abstract}
We show how to perform error correction of single qubit dephasing by
encoding a single qubit into a minimum of three. This may be performed 
in a manner closely analogous to classical error correction schemes. 
Further, the resulting quantum error correction schemes are trivially
generalized to the minimal encoding of arbitrarily many qubits so as to 
allow for multiqubit dephasing correction under the sole condition that 
the environment acts independently on each qubit.
\end{abstract}


The ability to store and manipulate quantum information for long 
periods is at the heart of many exciting new applications such as 
quantum computation
\cite{revs}, quantum communication across noisy channels \cite{Pure},
quantum cryptographic schemes \cite{Weis} and networks \cite{Biham} as 
well as possible attacks on simpler quantum cryptosystems \cite{aref}. 
The hope of realising such storage has been spurred by the 
theoretical construction of quantum error correction codes and circuits 
\cite{Shor,Shor2,Steane,Zurek,Ekert}. The original scheme corrected for 
arbitrary one-qubit errors in a single qubit encoded within an 
error-correcting coded state requiring 9 qubits \cite{Shor}. (A qubit is 
the information encoded in a two-state quantum system \cite{Sch}.)
This code
was soon reduced to requiring 7 qubits \cite{Steane} and finally the
minimum code requiring only 5 qubits \cite{Zurek}. The importance of using
the minimal resources in constructing quantum error correcting codes
is based on our current difficulty in performing operations on even two 
qubits \cite{Wineland,Kimble}. In this context, a scheme which could
perform quantum error correction with even fewer qubits could have 
important consequences. This paper shows that 1-qubit dephasing 
can be corrected with a minimum of 3 qubits encoding a single qubit
of quantum information. A similar scheme for correcting
dephasing was presented by Steane \cite{Steane2}. However, it uses 3 
auxiliary qubits for encoding  which is a total of 4 qubits. Futher, 
Steane's scheme uses external detection and manipulation conditioned
the measurement results for decoding. By contrast, the schemes discussed 
here work without the necessity of external detection.

To date, dephasing is the primary anticipated cause of failure
of a quantum computation \cite{Unruh,ShorZ,Bar} and of quantum 
information storage in general. In fact, it is the rule that
dephasing time for a quantum system is no longer, and usually a 
lot shorter, than the population decay time. With a suitable design, 
quantum computers might even approach the performance of classical 
computers for their insensitivity to random bit-flips. Indeed, classical 
computers virtually run without error correction \cite{fn} except over 
comparatively noisy networks.

We demonstrate here a recipe for converting classical
coding schemes, which protect against random bit-flips, to quantum 
schemes for protecting against dephasing. In the simplest case this
leads to a 3-qubit scheme of pure-quantum error correction.
More generally, it automatically provides us with the minimal 
encoding and decoding circuits to store arbitrary numbers of qubits 
and to correct against multibit dephasing. The generalization requires 
an environment that acts independently on each qubit (and through 
these interactions causes decoherence). Since between computational 
steps the qubits are decoupled one from another, it is expected that 
this is not too limiting a restriction on the applicability  of the
results presented here. In any case, the economy of the schemes 
make them attractive for implementation in at least the first 
generation of quantum memories and computers.

% FIG 1a
\begin{figure}[thb]
\begin{psfrags}
\psfrag{a}[r]{\LARGE a)}
\psfrag{b}[r]{\LARGE b)}
\psfrag{p}[c]{$\alpha|0\rangle+\beta|1\rangle{~~~~~}$}
\psfrag{p1}[l]{$\alpha\, e^{i\varphi_0}|0\rangle
              +\beta\, e^{i\varphi_1}|1\rangle$}
\psfrag{p2}[l]{$~~(e^{i\varphi_1}\!+e^{i\varphi_0})
                       (\alpha|0\rangle+\beta|1\rangle)$}
\psfrag{p3}[lu]{$+(e^{i\varphi_1}\!-e^{i\varphi_0})
                       (\alpha|1\rangle+\beta|0\rangle)$}
\psfrag{e}[c]{${~~~}$dephasing}
\psfrag{u1}[cb]{\small $~~~~\hat U$}
\psfrag{u2}[cb]{\small $~~~~\hat U^\dagger$}
\epsfxsize=3.4in
\epsfbox[-20 -20 270 160]{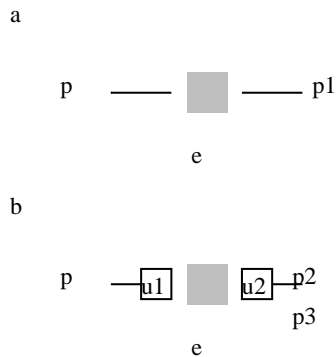}
\end{psfrags}
\caption{Action of dephasing on a single qubit. The qubits pass
from left to right in these `circuits.' When the qubit passes a) through
the shaded region it experiences random dephasings (this
region represents the coupling to an environment); b) through this 
region, between the sequence $\hat U$, $\hat U^\dagger$, it experiences
random bit-flipping. [Here $\protect\hat U$ is the rotation
$\protect\exp(-i\protect\pi\protect\hat\protect\sigma_y/4)$ and the
outgoing states are not normalized.]}
\label{fig1a}
\end{figure}

The schemes studied here rely on the following geometric representation:
Single qubit dephasing is generated by random rotations about the 
$z$-axis of the Bloch sphere. Such single particle dephasings may be
individually converted into random bit-flips by a $\pi/2$ rotation
about the $y$-axis of the Bloch sphere of each particle, see
Fig.~\ref{fig1a}. Random bit flips alone, however, can be optimally 
corrected by classical coding schemes \cite{abook}. Thus, by taking 
a classical error correction circuit and reinterpreting it as acting 
on qubits, we may use it to correct independent qubit dephasing. 
In effect we are translating qubit dephasing into qubit flipping
for which the error-correction circuit can correct.

% FIG 1
\begin{figure}[thb]
\begin{psfrags}
\psfrag{a}[c]{\Large $\psi$}
\psfrag{z}[c]{\Large $0$}
\psfrag{e}[c]{${~~}$1-bit error}
\epsfxsize=3.4in
\epsfbox[-30 -20 230 115]{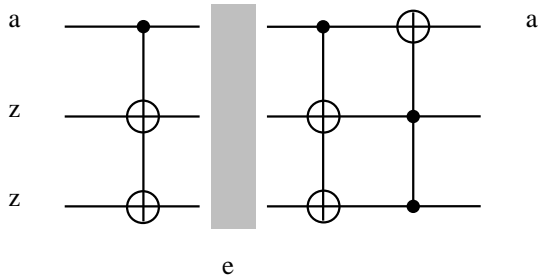}
\end{psfrags}
\caption{Classical error correction circuit. Here each classical
bit $\psi$ is recorded as a redundant triple $\psi\psi\psi$. The shaded
region represents the possible introduction of a single randomly flipped
bit. After this error the decoding circuit successfully restores the
original value of $\psi$ in the upper line.}
\label{fig1}
\end{figure}

Fig.~\ref{fig1} shows a classical circuit for encoding and decoding of
the classical bit $\psi\in \{0,\,1\}$ into three bits in order to 
protect it from single bit-flips. The bit $\psi$ enters the circuit 
from the left and `propagates' along the horizontal lines; the other 
two bits are initially $0$. The vertical lines represent various 
logical operations among the bits. The $\oplus$ symbols represent 
NOT operations, which by themselves would unconditionally flip bits. 
As they appear in Fig.~\ref{fig1}, however, they are connected
vertically to one or more bits having a black spot at the connection. 
In this case, the NOT's are conditional on all vertically connected 
black-spotted bits having the value $1$ --- these are conditional 
NOT gates. In Fig.~\ref{fig1} the encoding is performed by the first 
vertical line representing a conditioned double NOT. This encodes a 
$0$ in the upper bit as $000$ and a $1$ as $111$. Here the shaded
region in the circuit represents a classical one-bit flipping error. 
Including the possibility of no error occurring such a bit flip
could take the encoded $000$ to any of
$\{\,000, \, 001,\,010, \, 100\,\}$. Similarly the encoded $111$ could
be taken to any of $\{\,111, \, 110,\,101, \, 011\,\}$. The remainder 
of the circuit in Fig.~\ref{fig1} to the right of the shaded region 
performs the decoding. It simply reverses the coding sequence followed 
by a double conditioned NOT to correct for the case where the upper 
bit was flipped in the shaded region. Indeed, it is always the case 
that the encoding could be performed by running the decoding stage 
backwards. However, it is not necessarily the most efficient method. 

% One possible advantage
% of performing the encoding by the backwards run decoding scheme is
% that one might be able to use so-called ``cheap'' logic gates that
% introduce extraneous phases. Such cheap gates are easier to construct
% experimentally. To this author, at least, it is not yet obvious that
% such cheap solutions always work when sandwiched 
% around interactions with the environment.
% \notes{this para could be made simpler.  i still don't know how.}

\begin{figure}[thb]
\begin{psfrags}
\psfrag{a}[c]{\Large $|\psi\rangle$}
\psfrag{z}[c]{\Large $|0\rangle$}
\psfrag{e}[c]{${~~~}$1-bit dephasing}
\psfrag{u1}[cb]{\small $~~~~\hat U$}
\psfrag{u2}[cb]{\small $~~~~\hat U^\dagger$}
\epsfxsize=3.4in
\epsfbox[-30 -20 260 115]{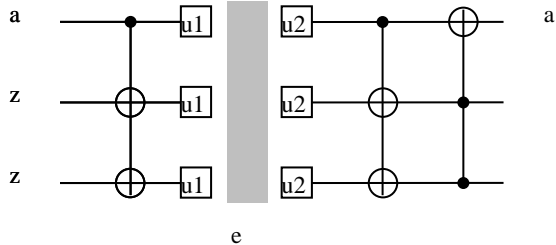}
\end{psfrags}
\caption{Quantum 1-bit dephasing correction. Here
$\protect\hat U$ is the rotation
$\protect\exp(-i\protect\pi\protect\hat\protect\sigma_y/4)$.}
\label{fig2}
\end{figure}

How do we take the classical circuit in Fig.~\ref{fig1} and convert 
it into a quantum error correction circuit for dephasing? Firstly, 
we interpret each of the elements quantum mechanically: the encoded 
bit is replaced by a qubit (any superposition of the basis elements 
$|0\rangle$ or $|1\rangle$); and the logical gates are replaced with 
quantum gates which may be built up in an elementary way \cite{elgates},
with logical operations applying to each branch of the wavefunction.
The final step is a $\pi/2$ rotation of each bit about the $y$-axis 
in the Bloch sphere. This completes the encoding. Now, when dephasing
occurs in a single qubit, we decode this error by first reversing the
rotation on each bit and then proceeding with the usual single 
classical-bit-flip circuitry, see Fig.~\ref{fig2}.

% FIG 2

The minimal requirements to correct one-bit 
dephasing of one encoded
qubit can be studied by arguments similar to those developed in
Ref.~\onlinecite{Zurek}. There the environment could produce `rotations'
in the Bloch sphere about any axis, whereas here we assume the environment
is free only to produce rotations about a single axis. Such arguments
show that the minimal dephasing correction scheme requires 3 qubits, 
which has just been achieved in Fig.~\ref{fig2}. More directly, however,
the circuit is optimal because the classical version of it was.

% FIG 3
\begin{figure}[thb]
\begin{psfrags}
\psfrag{p1}[c]{\Large ${~~}|\psi\rangle$}
\psfrag{p2}[cr]{\Large ${~}|\psi\rangle$}
\psfrag{z}[c]{\Large $|0\rangle$}
\psfrag{u1}[cb]{$~~~~\hat U$}
\psfrag{u2}[cb]{$~~~~\,\hat U^{\!\dagger}$}
\psfrag{e}[b]{${~~~~}$2-bit dephasing}
\psfrag{m}[c][b][1][90]{\small {$~~\,$}majority}
\epsfxsize=3.4in
\epsfbox[-20 -10 215 210]{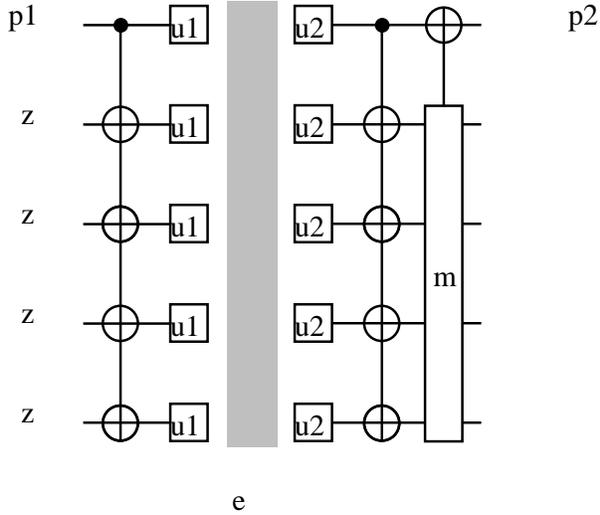}
\end{psfrags}
\caption{Quantum 2-bit dephasing correction of the qubit 
$|\protect\psi\protect\rangle$ requiring a minimum of 5 qubits. The
decoding requires a new gate which is a bit flip $\oplus$ conditioned
on the majority of the qubits passing through the `majority' box
being in state $|1\protect\rangle$. Such multibit dephasing generalizations 
require that the environment dephases each qubit independently.}
\label{fig3}
\end{figure}

Generalizing the scheme shown in Fig.~\ref{fig2} is also trivial.
Taking a classical coding scheme, we perform single bit rotations
to convert the single bit dephasings into bit flips. Examples of such
generalizations are easily constructed as shown in Fig.~\ref{fig3} for
a scheme protecting one qubit against two-bit dephasing. Again, it
is easy to check that this quantum scheme is minimal for what it achieves.
The main disadvantage of these generalizations is not immediately
obvious from our discussion of dephasing or quantum error correction:
The scheme is designed to correct dephasing which was incurred 
independently on each qubit (by independent interactions with the 
environment). If, however, a conditional dephasing occurs on one qubit 
based on the state of a second qubit then our scheme fails. Such errors
are {\it not\/} converted into bit-flip errors by our strategy and so 
the `classical' schemes employed are inadequate.

Two final issues require addressing in {\it all\/} quantum error 
correction schemes. First,  the circuitry of quantum error correction 
schemes are typically derived and work when `expensive' gates 
\cite{elgates} are employed --- gates having no extraneous signs. 
Under purely unitary evolution the presence of extraneous signs is 
unimportant if one follows the simple rules of reversible programming 
\cite{elgates}. However, in quantum error correction, where part of 
the evolution is non-unitary, these rules are inadequate. {\it It 
appears that all quantum error correction circuits constructed so far 
require expensive gates.}

Second, error correction is, in some sense, 
static; it aims at `refrigerating' particular degrees of freedom 
of interest to us and is ideal for providing us with stable quantum 
memories. But how do we combine this static nature with the dynamics 
necessary for computation? This is an unsolved problem. Computational 
(dynamical) steps are indistinguishable from many errors. Therefore, 
in order to compute we must make the computer susceptible to errors, 
by suspending error correction. One proposed solution \cite{Steane2} is
to apply error correction to the overall state of the computer between 
computational steps. In this way the correction occurs only while 
the computer is static. Unfortunately, this also means that all 
errors during the actual computation remain uncorrected and accumulate 
with every step taken. This application yields no advantage over an 
error correction free computation. 

An alternate approach is to apply error correction to those bits unused 
in the current computational step. Naturally, this alternative fails 
to protect against errors in bits that are involved in the computation; 
in principle, these could be as few as two bits. Yet another approach 
is to attempt to design computational steps mapping states protected by 
error correction code directly to each other. In this way we could 
`freeze' the state into the correct result of the computation using 
`dynamic error correction.' The last approach appears to be complicated. 
Clearly, more work is required.

In conclusion, we have shown how to perform error correction of 
single-qubit dephasing by encoding a single qubit into the minimum 
of three. This economy in qubits makes this scheme likely to be one of 
the first to be implemented. The scheme described is simple to 
construct, understand and generalize. Generalizations to multiqubit 
dephasing corrections, however, require that the environment acts 
independently on each qubit.

\vskip 0.25truein

The author appreciated the support of a Humboldt fellowship.


\begin{references}
\bibitem{fn}The fact that even the inclusion of parity check bits 
(the 9th bit in a byte of memory) occurs in only roughly half the 
commercial computers today suggests that random bit flips are rare.
\end{references}
\end{document}